\documentstyle[12pt,aasms4]{article}

\def\t0{\theta_{\circ}}

\def\be{\begin{equation}}
\def\en{\end{equation}}

\def\msun{M_{\sun}}

\def\lsun{L_{\sun}}

\begin{document}

\title {Mid-infrared imaging of the massive young star AFGL 2591:\\
Probing the circumstellar environment of an outflow source}
\author{Massimo Marengo \altaffilmark{1,2}, 
Ray Jayawardhana\altaffilmark{1,2,3},
Giovanni G. Fazio\altaffilmark{1,2},
William F. Hoffmann\altaffilmark{2,4},
Joseph L. Hora\altaffilmark{1},
Aditya Dayal\altaffilmark{5},
and Lynne K. Deutsch\altaffilmark{6}}
\altaffiltext{1}{Harvard-Smithsonian Center for Astrophysics, 60 Garden St., Cambridge, MA 02138; Electronic mail: mmarengo@cfa.harvard.edu}
\altaffiltext{2} {Visiting Astronomer, NASA Infrared Telescope Facility}
\altaffiltext{3} {Department of Astronomy, University of California at Berkeley, 601 Campbell Hall, Berkeley, CA 94720; Electronic mail: rayjay@astro.berkeley.edu}
\altaffiltext{4}{Steward Observatory, University of Arizona, Tucson, AZ  85721}
\altaffiltext{5}{IPAC/Caltech, MS 100-22, 770 S. Wilson Ave., Pasadena, CA 91125}
\altaffiltext{6}{Department of Astronomy, CAS 519, Boston University, 725 Commonwealth Ave., Boston, MA  02215}

\begin{abstract}
Most, if not all, stars are now believed to produce energetic outflows during
their formation. Yet, almost 20 years after the discovery of bipolar outflows 
from young stars, the origins of this violent phenomenon are not well
understood. One of the difficulties of probing the outflow process, 
particularly in the case of massive embedded stars, is a deficit of high 
spatial resolution observations. Here, we present sub-arcsecond-resolution 
mid-infrared images of one massive young stellar object, AFGL 2591, and
its immediate surroundings. Our images, at 11.7, 12.5 and 18.0 $\mu$m,
reveal a knot of emission $\approx$6'' SW of the star, which may be evidence 
for a recent ejection event or an embedded companion star. This knot is 
roughly coincident with a previously seen near-infrared reflection nebula 
and a radio source, and lies within the known large-scale CO outflow. We also 
find a new faint NW source which may be another embedded lower-luminosity 
star. The {\it IRAS} mid-infrared spectrum of AFGL 2591 shows a large silicate 
absorption feature at 10$\mu$m, implying that the primary source is surrounded 
by an optically thick dusty envelope. We discuss the interrelationship of 
these phenomena and suggest that mid-infrared imaging and spectroscopy provide
powerful tools for probing massive star birth.
\end{abstract}


\section{Introduction}
It is now generally agreed that most, if not all, stars produce energetic
outflows during their formation (see Bachiller 1996 for a review). Since the 
discovery of outflows almost 20 years ago (Snell, Loren \& Plambeck 
1980), nearly 200 examples have been found, most of them associated with
low-luminosity young stellar objects (YSOs). Outflows can be studied
using a variety of tracers including high-velocity CO line emission
(Bally \& Lada 1983), shock-excited H$_{2}$ line emission (e.g., Lane 1989), 
and optically visible Herbig-Haro (HH) jets (Reipurth 1989, 1999). 

All indications are that flows emerge bipolarly from a stellar or 
circumstellar region. However, the physical processes giving rise to the
outflows are not well understood. One suggestion is that a primary wind
originating at the star (e.g., Shu et al. 1991) or at the accretion disk 
(e.g., Pudritz \& Norman 1983) is responsible. In both scenarios, the fast 
well-collimated wind sweeps up ambient molecular gas in its vicinity, 
forming two cavities in opposite directions from the star. The displaced 
molecular gas constitutes the CO outflow and ionized ``knots'' of gas 
manifest themselves as HH objects. Outflows probably occur while the star 
is still accreting material, and may play a significant role in ejecting 
material of excess angular momentum. 

The presence of ``bullets'' in the CO emission and shocked ``knots'' visible
in the optical and near-infrared suggests that mass ejection from YSOs
often occurs in bursts rather than at a steady rate. Such ejection events
appear to be intermittent, and the interval between two successive events
is on the order of $10^3$ years. It is intriguing that the masses and 
timescales of bullets are similar to those of outbursts observed in FU Ori
stars. The FU Ori eruptions are thought to be due to large increases in
the accretion rate through a circumstellar disk (Hartmann \& Kenyon 1985).
If the two phenomena are related, it may be possible to probe the
accretion history of protostars using detailed observations of their outflows
(e.g., Reipurth 1989).

One of the barriers to better understanding the outflow process is 
the limited spatial resolution of millimeter-wavelength CO line observations.
The situation is particularly dire in the case of massive young stellar 
objects in which large extinction ($A_{v} \gg$ 10 mag) prevents high-resolution
optical imaging. 

Here we present sub-arcsecond-resolution mid-infrared images of one 
high-luminosity (Class 0/I) YSO, AFGL 2591, and its immediate surroundings. 
AFGL 2591 was first recognized as an outflow source by Bally \& Lada (1983). 
The CO outflow is larger than 1' in extent, with the blue lobe extending west
or southwest and the red lobe in the northeast direction (Lada et al. 1984;
Mitchell, Hasegawa, \& Schella 1992; Hasegawa \& Mitchell 1995). 
Optical and near-infrared observations have revealed molecular hydrogen
emission (40'' east and west, Tamura \& Yamashita 1992) and HH objects
(10'' and 20'' west/southwest, Poetzel et al. 1992) near AFGL 2591. 
The distance to AFGL 2591 is not well established. Assuming a distance of 
1 kpc, it has a luminosity of $\sim 2 \times 10^4 \lsun$ and an estimated 
stellar mass of 10 $\msun$. 

\section{Observations and data reduction}
AFGL 2591 was observed on June 3 and 4, 1999 using MIRAC3, the
newest version of the University of Arizona/SAO mid-infrared camera
(Hoffmann et al. 1998) at the 3.0-meter NASA Infrared Telescope
Facility (IRTF). MIRAC3 uses a Boeing HF16 128$\times$128 Si:As
blocked impurity band detector. On the IRTF, MIRAC3 has a plate scale
of 0.33''/pixel, providing a total field of view of
42''$\times$42''. This pixel scale ensures Nyquist sampling of 
the diffraction-limited point spread function (PSF).

We obtained images of AFGL 2591 in MIRAC3 11.7, 12.5 and 18.0 $\mu$m 10\%
passband filters, with a total on-source integration time of 600, 370
and 540 seconds respectively. The standard stars $\alpha$ Her and
$\beta$ Peg, observed before and after the science target, were used
for flux and PSF calibration.

We used a standard nodding and chopping technique
to remove the background signal, dithering the source on the
array to obtain sub-pixel sampling of the PSF. 
The chop frequency was set to 5 Hz, with a throw of 20'' in the N-S
direction. The nod throw was also set to
20'' but in the E-W direction, in order to have all four chop-nod beams
inside the field of view of the array. Each individual nod cycle
required a 10-second on-source integration, and the procedure was
repeated for as many cycles as needed to obtain the requested total
integration time.

To reduce the data, we have developed our own software written in C
and IDL. We first subtracted the chop-on from the chop-off frames for 
both nodding beams. The two images thus obtained were then
substracted one with the other, in order to get a single frame in
which the source appears in all four beams (two negative and two
positive). We then applied a gain matrix, derived from images of the
dome (high intensity, uniform background) and the sky (low intensity,
uniform background), to flat field the chop-nodded image.

This procedure was repeated for each of the nodding cycles for 
which the source was observed. A final high S/N cumulative image was
then obtained coadding together all beams, each recentered and shifted 
on the source centroid. This last coadding was performed on a
sub-pixel grid having the size of one-fifth of the original MIRAC3
pixels, thus providing a final pixel scale of 0.066''/pixel. A mask
file to block out the effects of bad pixels and field vignetting was
also created and applied in this stage, preventing individual rejected 
pixels from contributing to the final image.

The same observing and reduction procedure was also used for the
reference stars, to ensure a uniform treatment of the source and the 
standards.

\section{Results and Discussion} 
Figure 1 shows the final MIRAC3 images of AFGL 2591. The central source 
is not clearly resolved. It has FWHM of 0.84'', 0.87'', and 1.21'' at 
11.7, 12.5 and 18.0 $\mu$m respectively. The FWHM of the reference
point source $\beta$ Peg in the same filters is 0.82'', 0.85'' and
1.11'' respectively. An extended knot of emission 
$\approx$6'' SW from  the central source is visible in all three images.
A compact faint source is seen $\approx$11'' NW of AFGL 2591. 
Figure 2 is a schematic drawing of AFGL 2591, not to scale,  with the
various observed components indicated.

In Table 1, we list mid-infrared fluxes for the three sources 
as well as the position angle of SW and NW sources (counter-clockwise
from North) and their separation from the central source. The photometry
was measured inside apertures of (diameter) 18'' (total flux of main and
SW sources), 9'' (SW source alone) and 5'' (NW source). The zero magnitude
flux densities for the three filters are 29.2, 25.7 and 12.6 Jy
respectively. We estimate an error of 10\% for the photometry, mainly
due to the uncertainty of the reference fluxes and the difficulty
of isolating the sources when doing aperture photometry. Lower fluxes
of the main source were found by Lada et al. (1984) by using broad-band
filters at 11.4 and 12.6 $\mu$m. The discrepancy (40\% and 8\%
in the two filters) can be explained by the presence of a broad absorption
feature in the source spectrum at 10 $\mu$m, which affects broad-band
photometry more than our 10\% filters. By fitting the 12.5 and 18.0 $\mu$m
fluxes with a black body, we derive color temperatures of 300-350K for the
central source, and 100-120K for the SW knot and the NW source. 

Persi et al. (1995) presented the only previously published mid-infrared 
image of AFGL 2591. Their observations at 11.7 $\mu$m were carried out 
on the 2.1-meter S. Pedro Martir telescope in Baja California, Mexico with 
an image scale of 1.34''/pixel, and did not reveal the SW knot or the NW 
compact source seen in our MIRAC3 images, due to the smaller field of
view of their detector.

\subsection {Central source}
The flux of the central source rises dramatically from 11.7 to 12.5 $\mu$m,
implying that its 11.7 $\mu$m flux may be suppressed by a large silicate 
absorption feature at $\sim$10$\mu$m. Indeed, the low-resolution mid-infrared 
spectrum obtained by the {\it InfraRed Astronomy Satellite (IRAS)} shows deep
absorption at that wavelength. Thus, it is likely that the protostar is 
surrounded by an optically thick dusty envelope, perhaps in free-fall 
collapse onto the star (van der Tak et al. 1999). Our observations limit 
the radius of the envelope to $\lesssim$3'' (or $\lesssim$ 3000 AU,
assuming a distance of 1~kpc).

Figure 3 shows the {\it IRAS} spectrum of the source, and the best fit we 
obtain by modeling the circumstellar dust envelope with the radiative 
transfer code DUSTY (Ivezi\'c et al, 1999). The fit shown is for a 
spherically symmetric envelope with a radial density profile 
$\rho \propto r^{-2}$ around a central black body source at 
effective temperature $T_{eff}$ = 25,000~K (van der Tek et al. 1999).
We adopted the silicate dust opacity by Ossenkopf et al. (1992), which
provides a good fit to the 10 $\mu$m silicate feature.
The fit for $\rho \propto r^{-1.5}$ (i.e., constant infall) is not as good, 
implying that the infalling envelope is likely to be surrounded by a fair 
amount of non-infalling cloud material. The model spectrum is not particularly
sensitive to the $T_{eff}$ of the central source. Our best fit to the 
{\it IRAS} spectrum suggests an $A_V$ on the order of 100 (derived
from the best fit optical depth).

\subsection {NW source}
AFGL 2591 is generally considered to be an example of relatively isolated 
massive star formation. However, our detection of a faint point-like
source 11'' NW of it suggests there may be other embedded low-mass stars
in its vicinity. A preliminary look at {\it Two-Micron All-Sky Survey (2MASS)}
J,H,K images reveals at least two more point-like sources within 40'' of 
AFGL 2591. Deep mid-infrared images of a larger area are needed to search for 
a possible cluster of low-mass protostars in this region.

\subsection {SW knot}
The SW knot we detect appears close to the inner portion of a
loop first observed in the near-infrared by Forrest \& Shure (1986).
The near-infrared emission is assumed to arise from scattering of photons 
from AFGL 2591 off dust grains at the surface of a cavity cleared by the 
outflow. The loop, also seen in NH$_3$ observations by Torrelles et al.
(1989), is indeed located in the blue outflow lobe, i.e., the one directed
toward us. Using Draine \& Lee (1984) opacity at 18$\mu$m and assuming 
$T=$100K and $d=$1~kpc, we estimate the mass of this knot to be 
$>$0.2$\msun$. 

The SW knot is also coincident with a radio source as well 
as a knot of H$_2$ emission. The radio source, detected using
interferometers at the Very Large Array (Campbell 1984) and 
the Owens Valley Radio Observatory (van der Tak et al. 1999), 
has an approximately flat spectrum between 6.1 and 0.3 cm, 
and is interpreted as free-free emission from an optically 
thin HII region. The H$_2$ emission is seen in an image taken
with the BEAR imaging spectrometer on the Canada-France-Hawaii 
telescope using a narrow band filter on the 1-0 S(1) line,
after the continuum is subtracted (G. Mitchell, personal 
communication). The existence of the H$_2$ line implies shock
collisional heating of the molecular gas in the outflow, which provide the
necessary excitation energy ($T \gtrsim 2000$~K, see e.g. Smith \& Brand,
1990). On the other hand, the dust responsible for mid-infrared emission 
is characterized by a much lower local equilibrium temperature of $T_d \sim
100$~K, as inferred from our photometry. It is likely that
the H$_2$ emission comes from the shock itself, while the mid-infrared
color temperature is an average through the dust clump.

Previous near-infrared observations have also revealed a second, larger 
loop of emission at a projected separation of $\sim$19'' from the star (Burns 
et al. 1989; Minchin et al. 1991; Tamura et al. 1991; Poetzel, Mundt, \& Ray 
1992). The presence of multiple shells/loops is thought to be evidence
for episodic mass outflows from AFGL 2591, as are molecular bullets seen in 
CO emission (Bachiller et al. 1991) and shocked knots observed in the optical 
and the near-infrared (Reipurth 1989) in other well-known outflow sources.
Up to $\sim 1 \msun$ could be ejected in each eruptive event, though a few 
tenths of a solar mass is more typical (Bachiller 1996).

If we assume that the SW knot is associated with a recent mass ejection from 
AFGL 2591, we can estimate its dynamical timescale. Following Minchin et al. 
(1991), we take that the outflow is inclined from the plane of the sky by 55 
degrees, and estimate $\sim 10^4$ AU as the true distance from the star to 
the SW knot. Given a true outflow velocity of 26 kms$^{-1}$ (based on the 
line-of-sight velocity of CO gas observed by Bally \& Lada 1983), then we 
get a dynamical timescale of $\sim$2000 years. For the same assumptions, the 
outer loop seen in the near-infrared would have a dynamical timescale of 
$\sim$6000 years. Such timescales are consistent with the estimated 
$10^4$--$10^5$-year total duration of the energetic outflow phase of YSOs 
(Bally \& Lada 1983).

Another possibility is that the SW source itself is an embedded protostar
surrounded by an ultracompact HII region. Some Herbig Ae/Be stars
--LkH$\alpha$ 198 and LkH$\alpha$ 234, for example-- are known to have 
heavily embedded mid-infrared companions associated with an outflow, a radio 
source, and HH objects (Lagage et al. 1993; Cabrit et al. 1997). 
The SW source is well-resolved in our mid-infrared images, with a
''coma"-like shape. So, if it is another protostar, its envelope may be
affected by the outflow from AFGL 2591. 

\smallskip
In summary, our observations suggest that mid-infrared imaging and 
spectroscopy can be powerful tools for probing the immediate circumstellar
environment of embedded stars driving outflows. Such outflows appear to 
be rather common around young stars and are very likely associated with 
accretion processes from the early stages of protostars. They may play an 
important role in determining the mass of the star-disk system, and affect 
the evolution of the surrounding molecular cloud. Yet our understanding of 
the outflow process is far from satisfactory. High-resolution imaging with 
current mid-infrared cameras and future millimeter-wave interferometers could 
help develop a more unified picture of infall and outflow in star formation.

\bigskip
We are grateful to George Mitchell, Charles Lada and Paolo Persi for
helpful discussions and to an anonymous referee for sensible suggestions.
We thank the IRTF staff for their outstanding support. The research 
was supported by the Smithsonian Institution and NASA grants to the
SIRTF-IRAC project. 

\newpage
\begin{table}
\begin{scriptsize}
\begin{center}
\renewcommand{\arraystretch}{1.2}
\begin{tabular}{lccccc}
\multicolumn{4}{c}{\scriptsize TABLE 1}\\
\multicolumn{4}{c}{\scriptsize MID-INFRARED PHOTOMETRY}\\
\hline
\hline
Source & 11.7$\mu$m (Jy) & 12.5$\mu$m (Jy) & 18.0$\mu$m (Jy) & PA & Separation\\
\hline
\hline
A & 440 ($\pm$44) & 745 ($\pm$75) & 753 ($\pm$75) & --- & --- \\
B & 18.4 ($\pm$2) & 15.9 ($\pm$2) & 79.6 ($\pm$8) &$235^\circ$ & 6''\\
C & 1.8 ($\pm$0.2) & 2.1 ($\pm$0.2) & 16.2 ($\pm$2) & $325^\circ$ & 11''\\
\hline
\end{tabular}
\end{center}
\end{scriptsize}
\end{table}

\newpage

\centerline{\bf Figure Captions}

\bigskip
\bigskip

Figure 1. MIRAC3 images of AFGL 2591 at (a) 11.7$\mu$m, (b) 12.5$\mu$m, and 
(c) 18.0$\mu$m. Contours are plotted for 10 and 15 $\sigma$ levels in
panels (a) and (b), and for 5, 10 and 15 $\sigma$ in panel (c).

Figure 2. A schematic drawing of AFGL 2591 with the various observed components
indicated. (Not to scale.)

Figure 3. {\it IRAS} mid-infrared spectrum of AFGL 2591 ({\it solid line})
and the best-fit model ({\it dashed line}) for the 10$\mu$m silicate
feature. The heavy absorption is presumably due to an optically thick
dusty envelope around the massive central star. 

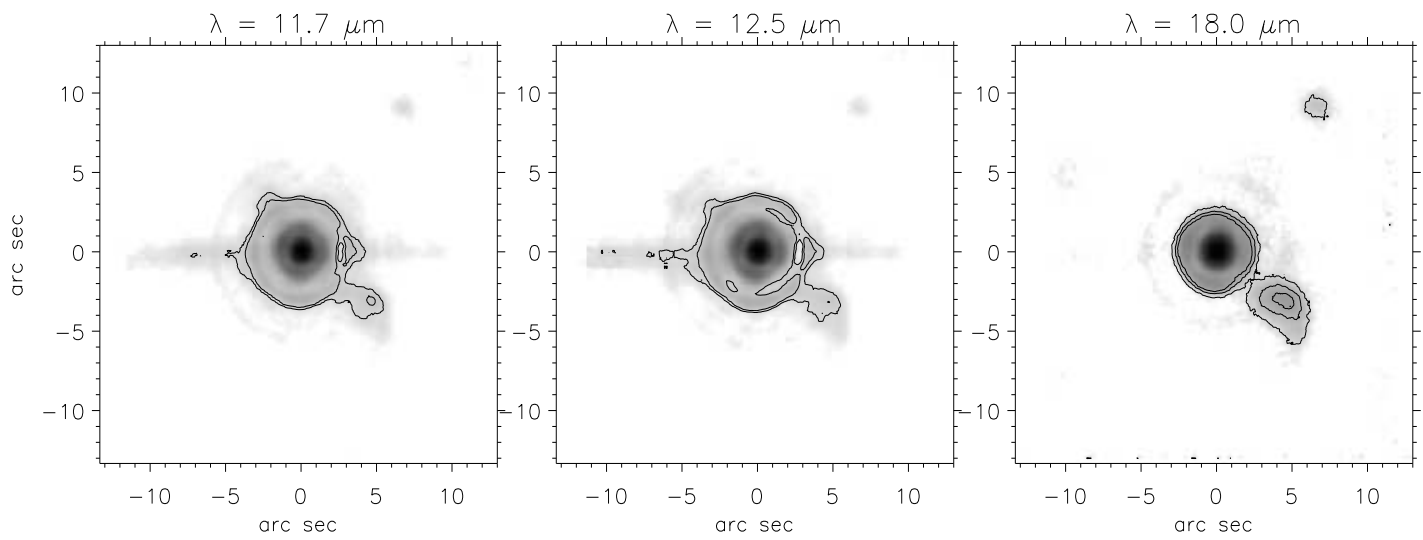
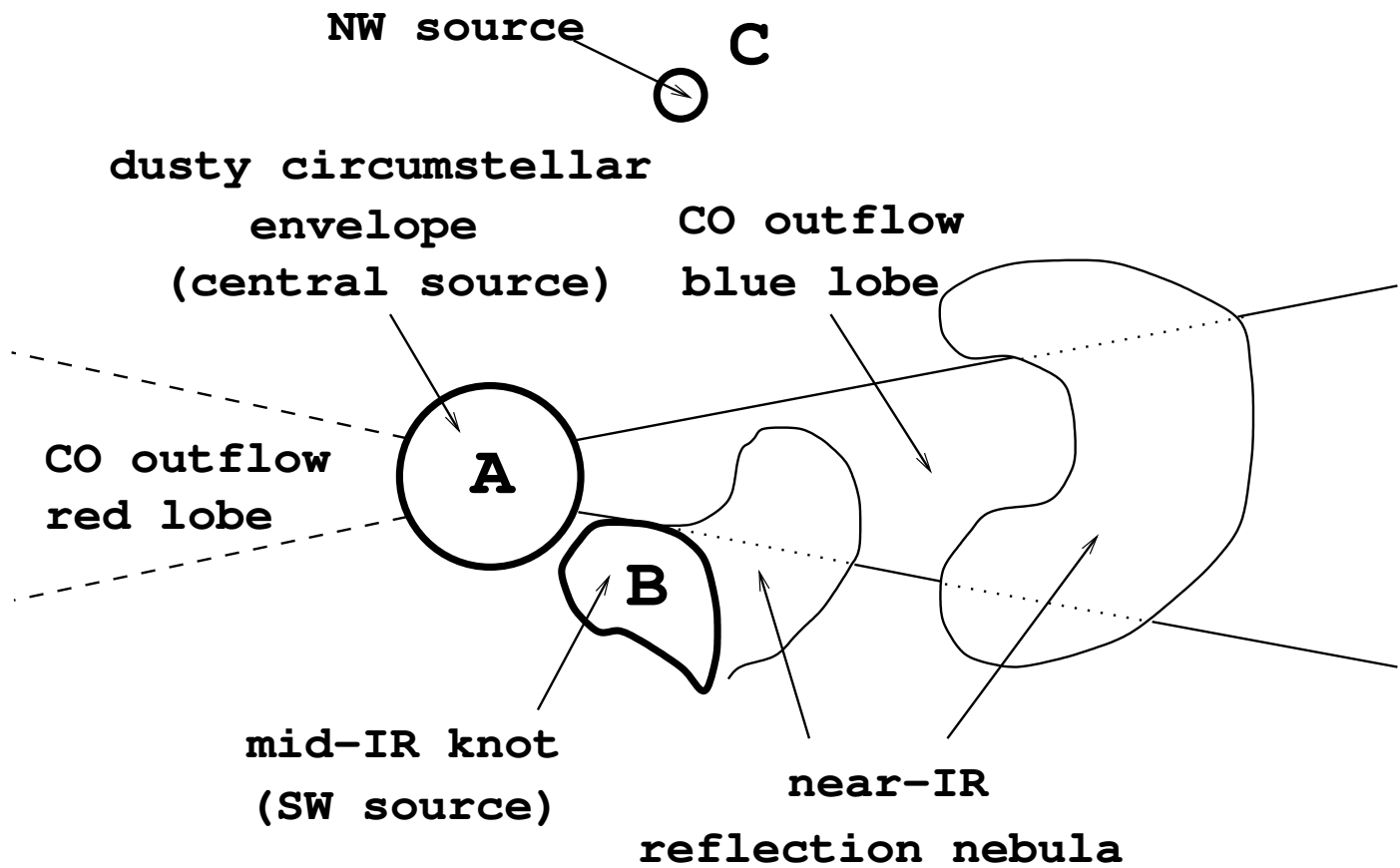
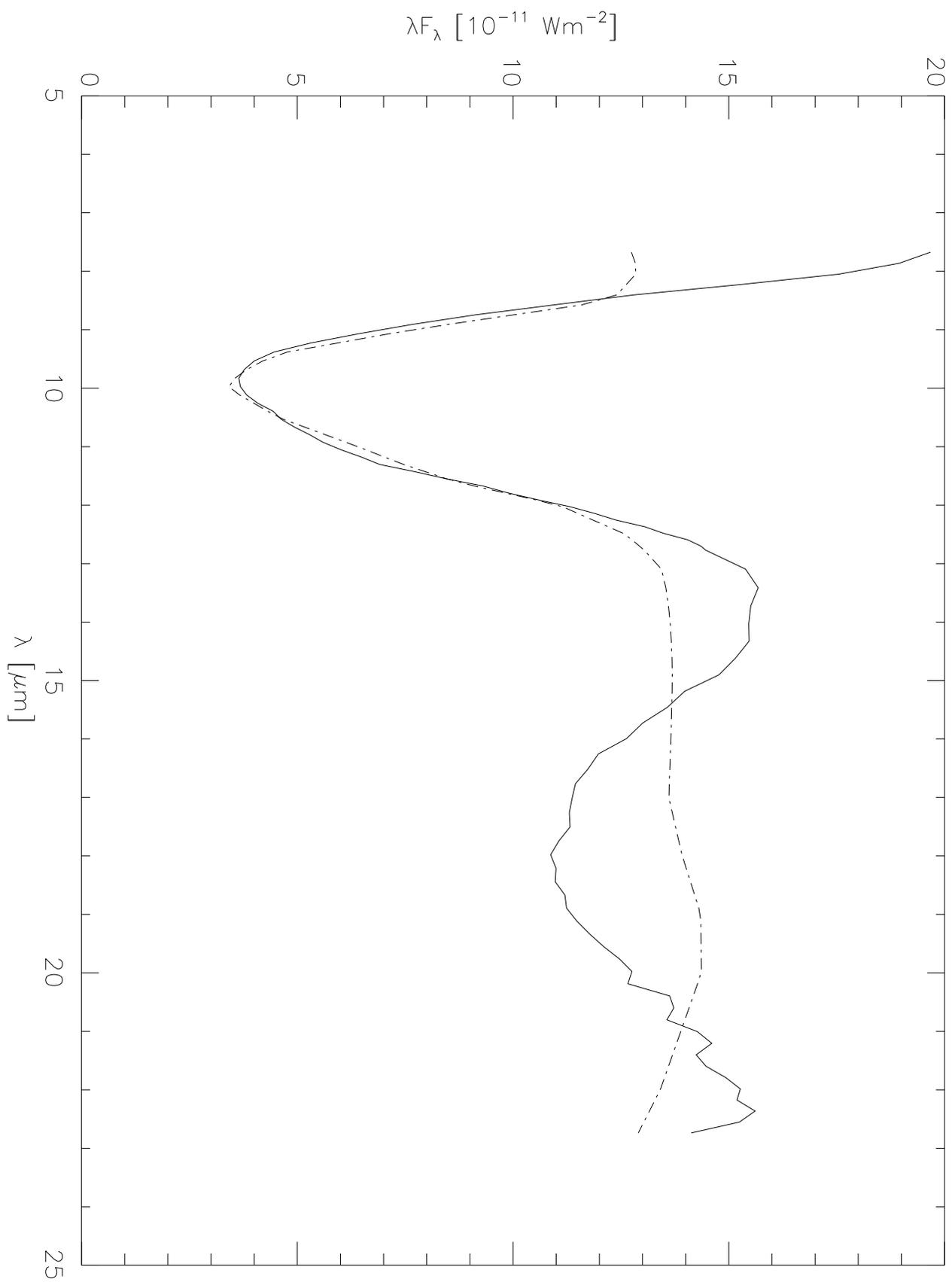
\end{document}